\documentclass[usenatbib]{mn2e}

\usepackage{graphicx}	
\usepackage{amsmath}	
\usepackage{amssymb}	

\newcommand{\be}{\begin{equation}}
\newcommand{\ee}{\end{equation}}
\newcommand{\bea}{\begin{eqnarray}}
\newcommand{\eea}{\end{eqnarray}}
\newcommand{\beas}{\begin{eqnarray*}}
\newcommand{\eeas}{\end{eqnarray*}}

\title[Universal Kepler Solver]{A Fast and Accurate Universal Kepler Solver without Stumpff Series}

\author[J.~Wisdom and D.M.~Hernandez]{
Jack Wisdom,$^{1}$\thanks{E-mail: wisdom@mit.edu (JW)}
and David M. Hernandez$^{1}$
\\
$^{1}$Massachusetts Institute of Technology, Cambridge, MA, 02139, USA
}

\date{Accepted XXX. Received YYY; in original form ZZZ}

\pubyear{2015}

\begin{document}
\label{firstpage}
\pagerange{\pageref{firstpage}--\pageref{lastpage}}
\maketitle

\begin{abstract}
We derive and present a fast and accurate solution of the initial
value problem for Keplerian motion in universal variables that does
not use the Stumpff series.  We find that it performs better than 
methods based on the Stumpff series.
\end{abstract}




\section{Introduction}
\label{intro}

\citet{WH91} introduced a symplectic mapping method for
the rapid simulation of the $n$-planet problem ($n$ planets plus a
massive central body).  The method splits the Hamiltonian for the
$n$-planet problem into Kepler Hamiltonians and an interaction
Hamiltonian, each of which may be efficiently solved.  The evolution
of the $n$-planet problem is obtained by interleaving the elementary
pieces.  The rapid and accurate solution of the Kepler initial value
problem is an essential part of the method.  The Wisdom-Holman method
and its variations has been widely adopted for solar system dynamics
investigations.

The method of \cite{WH91} evolved from the mapping method
of \citet{W82}; it relies on the averaging principle to introduce
Dirac delta functions into the Hamiltonian.  An alternate approach,
which leads to the same algorithm, is that of symplectic integration,
which uses the algebra of Lie series to approximate the local
evolution to some order in the stepsize by interleaving the evolution
of the pieces.  An advantage of the mapping approach is that the
stability of the method can be analyzed in terms of the overlap of
``stepsize resonances,'' which can be read off the delta function
Hamiltonian \citep{WH92}.  Another advantage of the
mapping approach is that perturbation theory can be used to improve
the method by eliminating the high-frequency terms introduced by the
delta functions, leading to the ``symplectic corrector'' \citep{WHT96}.  
An advantage of the symplectic integration
approach is its algebraic simplicity.

Jacobi coordinates were used in \cite{WH91} to eliminate the center of
mass freedom and to effect the separation of the Hamiltonian into
Keplerian and interaction parts.  \citet{TW93, TW94} used a
different splitting, making use of the canonical heliocentric
coordinates.  \citet{LD94} used the Wisdom-Holman method in Jacobi
coordinates, patched together with a non-symplectic method for
handling close encounters among the bodies.  They distributed their
program, called {\tt RMVS3}, in their {\tt SWIFT} package.
\citet{DLL98} used the canonical heliocentric variables with a
slightly different splitting; they call their method the ``democratic
heliocentric'' method.  They handle close encounters by recursively
subdividing the step in a symplectic manner.  Their program is
distributed as {\tt SYMBA}.  \citet{C99} introduced another method
based on the democratic heliocentric splitting, with a symplectic
transition to an ordinary numerical integration method (the
Bulirsch-Stoer method) for close encounters.  His program is
distributed as {\tt MERCURY}.  \citet{LD00} noticed that methods based
on the democratic heliocentric splitting are unstable for large
eccentricities, and modified their {\tt SYMBA} program to numerically
integrate ``close encounters with the Sun.''  Their modified program
is called ``modified {\tt SYMBA}.''  An essential element of all of
these variations on the Wisdom-Holman method is that one must solve
the Kepler initial value problem: given the position and velocity at
one time, the task is to find the position and velocity at a different
time (displaced by a timestep that can be either positive or
negative).

A recent contribution using the Lie series approach is \citet{HB15}.
They developed a symplectic integrator for the collisional $n$-body
problem.  This method also relies on the rapid and accurate solution
of the Kepler initial value problem.

We have found that the solution of the initial value problem may be
efficiently and accurately carried out in universal variables.  The
resulting formulation and program work for all orbits, whether they
are elliptic, parabolic, or hyperbolic.  The traditional presentation
of univeral variables (e.g. \citet{D92}) makes use of Stumpff
functions and calculates them using their series representation.
Argument four-folding is used to bring the function argument into an
interval near zero so that the series converge rapidly enough.  However, all of
the functions can be represented in terms of ordinary trigonometric
functions. But, a straightforward computation using this
representation has unpleasantly large error, so the series and
argument four-folding appear to be necessary.  We have been able
to circumvent this problem by reexpressing the functions in a
numerically well defined way.  The builtin trigonometric functions are
fast and accurate, and our solution of the initial value problem
compares favorably with solutions based on the Stumpff series.

\section{Kepler Problem}

The Hamiltonian for the Kepler problem is:
\be
H(t, {\bf x}, {\bf p}) = \frac{p^2}{2m} - \frac{\mu}{r} ,
\ee
where $p$ is the magnitude of ${\bf p}$ and $r$ is the magnitude of
${\bf x}$.  The constants $\mu$ and $m$ depend on the context in which
the Kepler problem is found.  For example, if ${\bf x} = {\bf x}_2 -
{\bf x}_1$ is the relative coordinate in the two-body problem, then we
would be led to take $m$ to be the reduced mass $(1/m_1 + 1/m_2)^{-1}$
and $\mu = G m_1 m_2$.  In the $n$-body problem, the constants may be
different, depending on the formulation (e.g.~\citet{WH91}).

Hamilton's equations may be written as a second order system:
\be
\ddot{\bf x} = - \frac{k {\bf x}}{r^3},
\label{eq:eom}
\ee
where the ``Kepler constant'' $k = \mu/m$.  The evolution depends on the
constants only through this combination.  We use the dot notation
for derivative with respect to time.  

\section{Derivation of the Solution}

A basic reference for the solution of the Kepler initial value problem
is \citet{D92}.  We refer the reader to that presentation of
universal variables and the Stumpff series for background on the
series approach.  We follow some aspects of that presentation here.
Our derivation is self-contained.

We introduce a new independent variable $s$ satisfying
\be
\frac{ds}{dt} = \frac{1}{r} .
\ee
The equation of motion, Eq.~(\ref{eq:eom}), becomes
\be
{\bf x}'' - \frac{r'}{r} {\bf x}' + \frac{k}{r} {\bf x} = 0,
\label{eq:eom2}
\ee
where prime indicates differentiation with respect to $s$.

The Hamiltonian is conserved, since there is no explicit time dependence.
It is conventional to write the conserved quantity as
\be
\beta = \frac{k}{a} = \frac{2 k}{r} - \dot{\bf x} \cdot \dot{\bf x} .
\ee
The constant $\beta$ is positive for elliptic motion, for which $a$ is the semimajor axis,
$\beta$ is negative for hyperbolic motion, and $\beta$ is zero for parabolic motion.

Expressing $\beta$ in terms of derivatives with respect to $s$ yields:
\be
\beta = \frac{2 k}{r} - \frac{1}{r^2} {\bf x}' \cdot {\bf x}'.
\ee
Differentiating this expression with respect to $s$ and using the derivative of the
equation of motion, Eq.~(\ref{eq:eom2}), yields:
\be
{\bf x}''' + \beta {\bf x}' = 0.
\label{eq:eom3}
\ee

Next we introduce the $f$ and $g$ functions:
\bea
{\bf x} &=& f {\bf x}_0 + g {\bf v}_0 , \nonumber \\
{\bf v} &=& \dot{f} {\bf x}_0 + \dot{g} {\bf v}_0 ,
\label{eq:fgdef}
\eea
where ${\bf x}_0$ is the initial position, and ${\bf v}_0$ is the
initial time rate of change of position (the initial velocity).
Substituting this into Eq.~(\ref{eq:eom3}), we find
\bea
f''' + \beta f' &=& 0 \nonumber \\
g''' + \beta g' &=& 0 ,
\label{eq:eomfgprime}
\eea
using the independence of initial position and velocity.
The equations are satisfied by an offset simple harmonic oscillation in $s$.

To determine the initial values for the derivatives, we start with Eq.(\ref{eq:eom}), to find
\bea
\ddot{f} + \frac{k}{r^3} f &=& 0 \nonumber \\
\ddot{g} + \frac{k}{r^3} g &=& 0 .
\label{eq:eomfg}
\eea
From Eq.~(\ref{eq:fgdef}), the initial values satisfy $f_0 = \dot{g}_0 = 1$ and $\dot{f}_0 = g_0 = 0$.
Then
\bea
f'_0 &=& r_0 \dot{f}_0 = 0 \nonumber \\
g'_0 &=& r_0 \dot{g}_0 = r_0 \nonumber \\
f''_0 &=& r_0^2 \ddot{f}_0 + r_0 \dot{r}_0 \dot{f}_0 = -\frac{k}{r_0} \nonumber \\
g''_0 &=& r_0^2 \ddot{f}_0 + r_0 \dot{r}_0 \dot{f}_0 = r_0 \dot{r}_0 .
\eea

It is convenient to define solutions of Eq.~(\ref{eq:eomfgprime}) in terms of 
functions $G^\beta_i(s)$ satisfying
\be
G^\beta_i(s) = \frac{d}{ds} G^\beta_{i+1}(s) .
\ee
We start the ladder with 
\be
G^\beta_0(s) = \cos (\sqrt{\beta} s) ,
\ee
for $\beta > 0$,
and
\be
G^\beta_0(s) = \cosh (\sqrt{-\beta} s) ,
\ee
for $\beta < 0$.  
We can take
\be
G^\beta_1(s) = \frac{\sin(\sqrt{\beta} s)}{\sqrt{\beta}},
\ee
for $\beta > 0$, and
\be
G^\beta_1(s) = \frac{\sinh(\sqrt{-\beta} s)}{\sqrt{-\beta}},
\ee
for $\beta < 0$.
We can then take 
\be
G^\beta_2(s) = (1 - \cos (\sqrt{\beta} s))/\beta ,
\label{eq:G2+}
\ee
for $\beta > 0$,
and
\be
G^\beta_2(s) = (1 - \cosh (\sqrt{-\beta} s))/\beta,
\label{eq:G2-}
\ee
for $\beta < 0$.  Onward,
\bea
G^\beta_3(s) &=& (s - \sin (\sqrt{\beta} s)/\sqrt{\beta})/\beta \nonumber \\
              &=& (s - G^\beta_1(s))/\beta .
\eea
for $\beta > 0$, and
\bea
G^\beta_3(s) &=& (s - \sinh (\sqrt{-\beta})/\sqrt{-\beta})/\beta \nonumber \\
              &=& (s - G^\beta_1(s))/\beta ,
\eea
for $\beta < 0$.  We could go on, but this is all we need here.
We have chosen the constants so that $G^\beta_i(0) = 0$ for $i>0$.
Notice that for $i<3$ these functions satisfy
\be
(G^\beta_i(s))''' + \beta (G^\beta_i(s))' = 0.
\ee

Now we can use these functions to find solutions for $f$ and $g$.  Let
\be
f(s) = A_f G^\beta_1(s) + B_f G^\beta_2(s) + C_f .
\ee
The condition that $f(0) = 1$ implies $C_f = 1$.  Then we form the derivative
\be
f'(s) = A_f G^\beta_0(s) + B_f G^\beta_1(s).
\ee
The condition that $f'(0) = 0$ implies $A_f = 0$.  The next derivative is
\be
f''(s) = B_f G^\beta_0(s),
\ee
using the fact that $A_f = 0$.
The condition that $f''(0) = -k/r_0$ implies that $B_f = -k/r_0$.
Putting it together we find
\be
f(s) = 1 - (k/r_0) G^\beta_2(s) .
\ee

Similarly, we let
\be
g(s) = A_g G^\beta_1(s) + B_g G^\beta_2(s) + C_g .
\ee
We find
\be
g(s) = r_0 G^\beta_1(s) + r_0 \dot{r}_0 G^\beta_2(s) .
\ee

From these we find $\dot{f}$ and $\dot{g}$ by using the relation
$ds/dt = 1/r$ to find
\bea
\dot{f}(s) &=& -(k/(r r_0)) G^\beta_1(s) ,\nonumber \\
\dot{g}(s) &=& (r_0/r)( G^\beta_0(s) + \dot{r}_0 G^\beta_1(s)).
\label{eq:fgdot}
\eea
But we still have to find $s$!

The relation between $t$ and $s$ is
\be
h = t - t_0 = \int_0^s r(s) ds.
\ee
Let's express $r(s)$ in terms of $G^\beta_i(s)$.  First, 
\be
r'_0 = r_0 \dot{r}_0 = {\bf x}_0 \cdot \dot{\bf x}_0 .
\label{eq:rp0}
\ee
After a small reduction, we find
\be
r''_0 = k - r_0 \beta ,
\label{eq:rpp0}
\ee
and
\be
r'''_0 + \beta r'_0 = 0.
\ee
So $r(s)$ may be expressed as
\be
r(s) = r_0  +  r'_0 G^\beta_1(s) + r''_0 G^\beta_2(s) .
\ee
Using Eqs.~(\ref{eq:rp0}) and (\ref{eq:rpp0})
\be
r(s) = r_0 G^\beta_0(s) + r_0 \dot{r}_0 G^\beta_1(s) + k G^\beta_2(s) .
\label{eq:r}
\ee
The properties of the $G^\beta_i$ allow an immediate
integration to find
\be
h = r_0 G^\beta_1(s) + r_0 \dot{r}_0 G^\beta_2(s) + k G^\beta_3(s) .
\label{eq:kepeqn}
\ee
This implicit equation for $s$ is our ``Kepler equation.''  We can
solve this by any of the standard methods: Newton, Halley, Laguerre,
bisection, and so on.  In practice, we first try Newton's method.  If
this fails, we try the Laguerre-Conway method. If this fails, we
recursively subdivide the step.

For the hyperbolic case our solution is as follows.  For small
stepsizes the equation is well approximated by a cubic equation.  We
take the real solution of this cubic equation as our initial guess in
Newton's method.  If Newton's method does not converge, then we use
the Laguerre-Conway method.  If this fails we recursively subdivide
the step.

Eq.~(\ref{eq:r}) can be used to rewrite the
equation for $\dot{g}$.  We find
\be
\dot{g}(s) = 1 - (k/r) G^\beta_2(s) .
\ee

The expressions for the parabolic case can be found by taking a limit
as $\beta$ goes to zero.  As it turns out the Kepler equation becomes
a cubic equation in $s$ and so can be solved without iteration.

\section{Numerical Refinement}

In order to have expressions that are well defined numerically we have
to do a little more work.

Examining Eqs.~(\ref{eq:G2+}) and (\ref{eq:G2-}), we see
that for small arguments of the cosine and hyperbolic cosine functions 
there will be cancellation and loss of precision.  But this problem can
be fixed by using the half-angle formulas.

Assuming $\beta>0$, let 
\bea
s_2 &=& \sin(\sqrt{\beta} s/2) \nonumber \\
c_2 &=& \cos(\sqrt{\beta} s/2) ,
\eea
then we can write
\bea
G^\beta_1(s) &=& 2 s_2 c_2/\sqrt{\beta} \nonumber \\
G^\beta_2(s) &=& 2 s_2 s_2/\beta \nonumber \\
G^\beta_3(s) &=& (s - G^\beta_1(s))/\beta \nonumber \\
G^\beta_0(s) &=& 1 - \beta G^\beta_2(s) .
\eea
The only expression that might be of concern is the expression for
$G^\beta_3(s)$, but, in practice, it seems to not be a problem.
Similar expressions can be derived for the $\beta < 0$ case.
Such changes are made throughout the program.

\section{Numerical Exploration}

We compare here our method and program ({\tt universal.c}), to two
universal variable Kepler stepper programs that use Stumpff series.
The program {\tt drift\_one.f} was written by Harold Levison and
Martin Duncan.  It is based on the presentation in \citet{D92}.  The
program {\tt drift\_one.f} is available as part of the {\tt SWIFT}
package.  This program is used in a number of programs derived from
the Wisdom-Holman method \citep{WH91}.  These programs
include {\tt RMVS3} \citep{LD94}, also distributed in
the {\tt SWIFT} package.  The same program is used to advance the
Kepler problem in the program {\tt SYMBA} \citep{DLL98}, and in {\tt MERCURY} \citep{C99}.  In order to compare to
our program, written in the C programming language, we have
translated, line by line, the program {\tt drift\_one.f} to a C
version {\tt drift\_one.c}.  We have {\em not} compared our code
directly to the fortran program {\tt drift\_one.f}.  We also compare
our program to the Kepler stepper in {\tt WHFast} \citep{RT15}.

Our test is designed to work for both elliptic and hyperbolic orbits.
Let $T = 2\pi / n$ where $n = \sqrt{k/|a|^3}$.  For elliptic orbits,
$T$ is the orbital period, and $n$ is the mean motion.  We choose $G =
(0.0172)^2$, which approximates the gravitational constant in units of
AU, day, and solar mass.  We take the mass factors to be one; thus the
Kepler constant $\mu/m$ is numerically just $G$.  We use the same
Kepler constant in the {\tt drift\_one.c} calculations.  We take the
semimajor axis for the elliptic case to be 0.4AU.  The eccentricity
and stepsize are varied.  The pericentric distance is $q = a(1-e)$,
and the velocity is determined from $a$ through the value of the
energy.  We evolve the Kepler orbit back and forth through pericenter,
adjusting the phase so that the pericenter is encountered with a wide
range of phases.  The chosen stepsize is $h$; we make use of the
auxillary stepsize $h' = \gamma h$, where $\gamma = (\sqrt{5} -1)/2
\approx 0.618$, is the irrational golden mean.  We start at pericenter
with $t=0$, and evolve the orbit with stepsize $h$ until $t>T/2$, i.e.
until we have passed apocenter (for elliptic orbits).  Then we adjust
the phase with a positive step of $h'$.  At this point we start
collecting statistics.  We reverse the timestep and evolve the orbit
with timestep $-h$ until $t<-T/2$.  Then we adjust the phase with a
positive step of $h'$.  We evolve with stepsize $h$ until $t>T/2$; we
adjust the phase with a postive step of $h'$, and so on.  We repeat
this process for 100 pericenter passages.  At the end we collect
statistics again, and compare to the initial statistics.  This
procedure tests all parts of the evolution for a large variety of
phases. It also tests both positive and negative timesteps.

The results for elliptic orbits for three methods are summarized in
Figs.~\ref{fig:fig1}-\ref{fig:fig3}.  The top plot in each set shows
the energy error as a function of eccentricity $e$ and stepsize $h$.
The middle plot shows the sign of the energy error at the end of each
evolution.  We would like this plot to show an intimate mixture of
positive and negative results, so that an evolution computed with the
Kepler solver does not show any tendency to expand or contract.  This
``bias'' plot was introduced by \citet{RT15}.  The bottom
plot shows an estimate of the computing time per Kepler step.  The
tests were run on a 4 GHz iMac, with an Intel i7 chip.  All of
the tests were compiled with C compiler optimizer option {\tt -O3}.
Comparing Figs.~\ref{fig:fig1} and \ref{fig:fig2}, we see that our
code is more accurate, faster, and shows less bias than {\tt
drift\_one.c}.  Note the complicated structure in the bias plot for
{\tt drift\_one.c}, and the intimate mixture of colors in the bias
plots for {\tt universal.c} and the Kepler stepper in {\tt WHFast}.

\begin{figure}
\begingroup%
\makeatletter%
\newcommand{\GNUPLOTspecial}{%
  \@sanitize\catcode`\%=14\relax\special}%
\setlength{\unitlength}{0.0500bp}%
\begin{picture}(4680,6048)(0,0)%
  \special{psfile=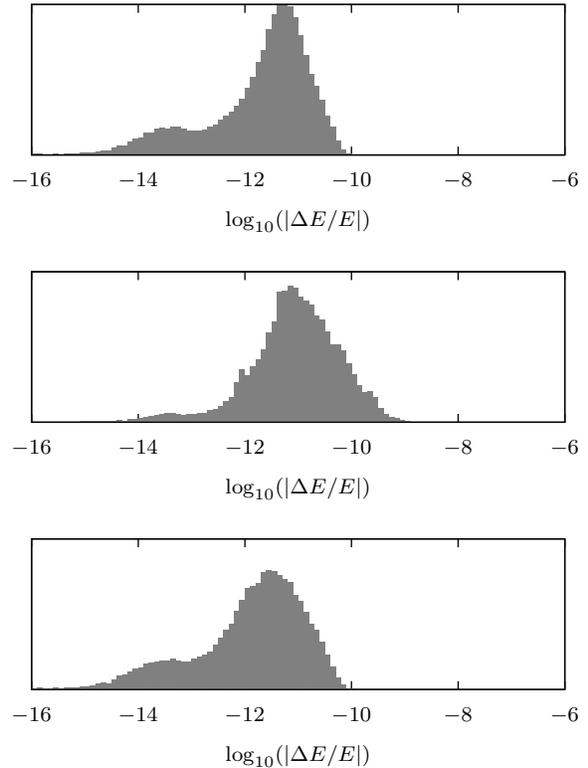 llx=0 lly=0 urx=234 ury=302 rwi=2340}
  \put(2309,4172){\makebox(0,0){\strut{}$\log_{10}(|\Delta E/E|)$}}%
  \put(4319,4472){\makebox(0,0){\strut{}$-6$}}%
  \put(3515,4472){\makebox(0,0){\strut{}$-8$}}%
  \put(2711,4472){\makebox(0,0){\strut{}$-10$}}%
  \put(1908,4472){\makebox(0,0){\strut{}$-12$}}%
  \put(1104,4472){\makebox(0,0){\strut{}$-14$}}%
  \put(300,4472){\makebox(0,0){\strut{}$-16$}}%
  \put(2309,2156){\makebox(0,0){\strut{}$\log_{10}(|\Delta E/E|)$}}%
  \put(4319,2456){\makebox(0,0){\strut{}$-6$}}%
  \put(3515,2456){\makebox(0,0){\strut{}$-8$}}%
  \put(2711,2456){\makebox(0,0){\strut{}$-10$}}%
  \put(1908,2456){\makebox(0,0){\strut{}$-12$}}%
  \put(1104,2456){\makebox(0,0){\strut{}$-14$}}%
  \put(300,2456){\makebox(0,0){\strut{}$-16$}}%
  \put(2309,140){\makebox(0,0){\strut{}$\log_{10}(|\Delta E/E|)$}}%
  \put(4319,440){\makebox(0,0){\strut{}$-6$}}%
  \put(3515,440){\makebox(0,0){\strut{}$-8$}}%
  \put(2711,440){\makebox(0,0){\strut{}$-10$}}%
  \put(1908,440){\makebox(0,0){\strut{}$-12$}}%
  \put(1104,440){\makebox(0,0){\strut{}$-14$}}%
  \put(300,440){\makebox(0,0){\strut{}$-16$}}%
\end{picture}%
\endgroup
 
\caption{These plots histogram the relative error in the elliptic case for three methods.
The bottom plot is for {\tt universal.c}, the middle plot is for 
{\tt drift\_one.c}, and the top plot is for the Kepler solver in {\tt WHFast}.}.
\label{fig:histogram-elliptic}
\end{figure}

We have computed the average ratio of the time per step for the
interval $0.001 < h/T < 0.1$, which is the range of most interest for
practical calculations, and find the average ratio for {\tt
drift\_one.c} relative to {\tt universal.c} is about 1.9.  The code
{\tt universal.c} is about twice as fast as {\tt drift\_one.c}.
We have also computed the average ratio of the time per step for the
interval $0.001 < h/T < 0.1$ for the Kepler solver in {\tt WHFast} and
find that the time per step is about a factor of 0.79 smaller than our
code---it is about $20\%$ faster.  In Fig.~\ref{fig:histogram-elliptic},
histograms of the relative energy error in the elliptic case for the
three methods are shown.  We see that {\tt drift\_one.c} is less
accurate than both {\tt universal.c} and the Kepler solver in {\tt
WHFast}.  Quantitatively, the averages of the common logarithm of the
absolute value of the relative error are: -11.92, -11.74, and -11.09,
for {\tt universal.c}, the Kepler solver in {\tt WHFast}, and {\tt drift\_one.c},
respectively.  The error of {\tt universal.c} is, on average, about
$50\%$ smaller than the Kepler solver in {\tt WHFast}, and about a
factor of 6.5 smaller than the error in {\tt drift\_one.c}.

For the hyperbolic case, we let $a = - 0.4AU$.  We start at
pericenter.  The pericentric distance is $q = a(1-e)$; $e$ is larger
than $1$.  The initial velocity is determined from the energy.  The
back and forth procedure is the same as in the elliptic case.  

The results for hyperbolic orbits for our code, {\tt universal.c}, and
for {\tt drift\_one.c} are shown in Figs.~\ref{fig:fig4} and
\ref{fig:fig5}.  \citet{RT15} did not extensively
test the hyperbolic case, and the Kepler solver in WHFast performs
poorly for unbound orbits.  Note again that the bias plot for our method {\tt
universal.c} shows an intimate mixture, whereas the bias plot for {\tt
drift\_one.c} shows evidence of bias.  We find that the average ratio
of the time per step of {\tt drift\_one.c} to the time per step of
{\tt universal.c} for the interval $0.001 < h/T < 0.1$ is about 1.6;
{\tt drift\_one.c} is about $60\%$ slower than {\tt universal.c}.  

\begin{figure}
\begingroup%
\makeatletter%
\newcommand{\GNUPLOTspecial}{%
  \@sanitize\catcode`\%=14\relax\special}%
\setlength{\unitlength}{0.0500bp}%
\begin{picture}(4680,4032)(0,0)%
  \special{psfile=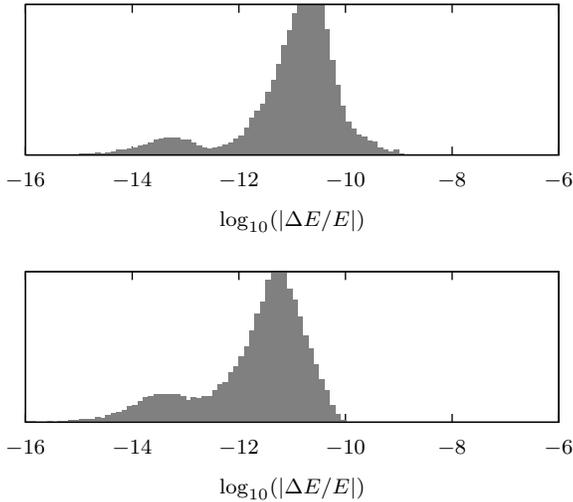 llx=0 lly=0 urx=234 ury=202 rwi=2340}
  \put(2309,140){\makebox(0,0){\strut{}$\log_{10}(|\Delta E/E|)$}}%
  \put(4319,440){\makebox(0,0){\strut{}$-6$}}%
  \put(3515,440){\makebox(0,0){\strut{}$-8$}}%
  \put(2711,440){\makebox(0,0){\strut{}$-10$}}%
  \put(1908,440){\makebox(0,0){\strut{}$-12$}}%
  \put(1104,440){\makebox(0,0){\strut{}$-14$}}%
  \put(300,440){\makebox(0,0){\strut{}$-16$}}%
  \put(2309,2156){\makebox(0,0){\strut{}$\log_{10}(|\Delta E/E|)$}}%
  \put(4319,2456){\makebox(0,0){\strut{}$-6$}}%
  \put(3515,2456){\makebox(0,0){\strut{}$-8$}}%
  \put(2711,2456){\makebox(0,0){\strut{}$-10$}}%
  \put(1908,2456){\makebox(0,0){\strut{}$-12$}}%
  \put(1104,2456){\makebox(0,0){\strut{}$-14$}}%
  \put(300,2456){\makebox(0,0){\strut{}$-16$}}%
\end{picture}%
\endgroup
 
\caption{These plots histogram the relative error in the hyperbolic case for two methods.
The bottom plot is for {\tt universal.c}, and the top plot is for 
{\tt drift\_one.c}.}
\label{fig:histogram-hyperbolic}
\end{figure}

\begin{figure*}
\begingroup%
\makeatletter%
\newcommand{\GNUPLOTspecial}{%
  \@sanitize\catcode`\%=14\relax\special}%
\setlength{\unitlength}{0.0500bp}%
\begin{picture}(7200,4536)(0,0)%
  \special{psfile=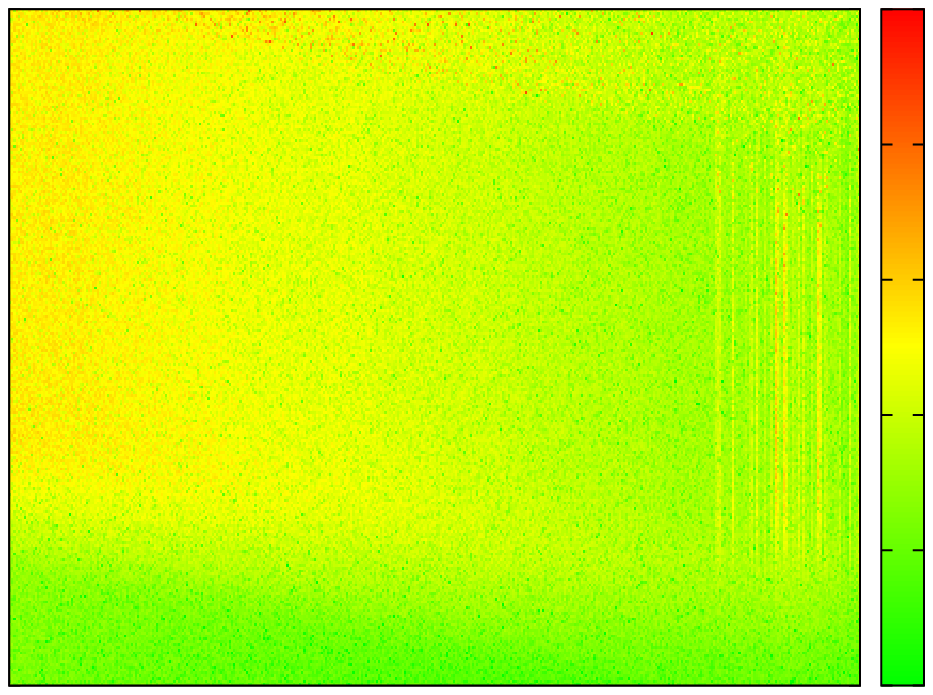 llx=0 lly=0 urx=360 ury=227 rwi=3600}
  \put(6369,4295){\makebox(0,0)[l]{\strut{}-6.0}}%
  \put(6369,3516){\makebox(0,0)[l]{\strut{}-8.0}}%
  \put(6369,2737){\makebox(0,0)[l]{\strut{}-10.0}}%
  \put(6369,1958){\makebox(0,0)[l]{\strut{}-12.0}}%
  \put(6369,1179){\makebox(0,0)[l]{\strut{}-14.0}}%
  \put(6369,400){\makebox(0,0)[l]{\strut{}-16.0}}%
  \put(160,2347){%
  \special{ps: gsave currentpoint currentpoint translate
630 rotate neg exch neg exch translate}%
  \makebox(0,0){\strut{}$\log_{10} (1 - e)$}%
  \special{ps: currentpoint grestore moveto}%
  }%
  \put(4249,200){\makebox(0,0){\strut{}$\ $}}%
  \put(2612,200){\makebox(0,0){\strut{}$\ $}}%
  \put(860,4295){\makebox(0,0)[r]{\strut{}$\mbox{\em}-8$}}%
  \put(860,3325){\makebox(0,0)[r]{\strut{}$\mbox{\em}-6$}}%
  \put(860,2348){\makebox(0,0)[r]{\strut{}$\mbox{\em}-4$}}%
  \put(860,1370){\makebox(0,0)[r]{\strut{}$\mbox{\em}-2$}}%
  \put(860,400){\makebox(0,0)[r]{\strut{}$\mbox{\em}0$}}%
\end{picture}%
\endgroup
 
\vskip -20pt
\begingroup%
\makeatletter%
\newcommand{\GNUPLOTspecial}{%
  \@sanitize\catcode`\%=14\relax\special}%
\setlength{\unitlength}{0.0500bp}%
\begin{picture}(7200,4536)(0,0)%
  \special{psfile=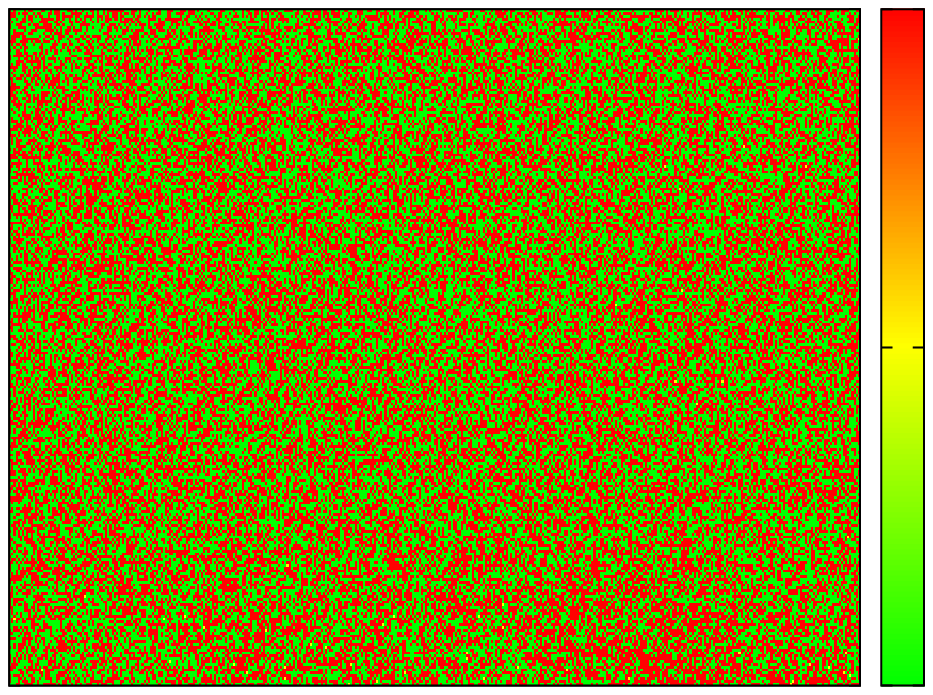 llx=0 lly=0 urx=360 ury=227 rwi=3600}
  \put(6369,4295){\makebox(0,0)[l]{\strut{}1.0}}%
  \put(6369,2347){\makebox(0,0)[l]{\strut{}0.0}}%
  \put(6369,400){\makebox(0,0)[l]{\strut{}-1.0}}%
  \put(160,2347){%
  \special{ps: gsave currentpoint currentpoint translate
630 rotate neg exch neg exch translate}%
  \makebox(0,0){\strut{}$\log_{10} (1 - e)$}%
  \special{ps: currentpoint grestore moveto}%
  }%
  \put(4249,200){\makebox(0,0){\strut{}$\ $}}%
  \put(2612,200){\makebox(0,0){\strut{}$\ $}}%
  \put(860,4295){\makebox(0,0)[r]{\strut{}$\mbox{\em}-8$}}%
  \put(860,3325){\makebox(0,0)[r]{\strut{}$\mbox{\em}-6$}}%
  \put(860,2348){\makebox(0,0)[r]{\strut{}$\mbox{\em}-4$}}%
  \put(860,1370){\makebox(0,0)[r]{\strut{}$\mbox{\em}-2$}}%
  \put(860,400){\makebox(0,0)[r]{\strut{}$\mbox{\em}0$}}%
\end{picture}%
\endgroup
 
\vskip -20pt
\begingroup%
\makeatletter%
\newcommand{\GNUPLOTspecial}{%
  \@sanitize\catcode`\%=14\relax\special}%
\setlength{\unitlength}{0.0500bp}%
\begin{picture}(7200,4536)(0,0)%
  \special{psfile=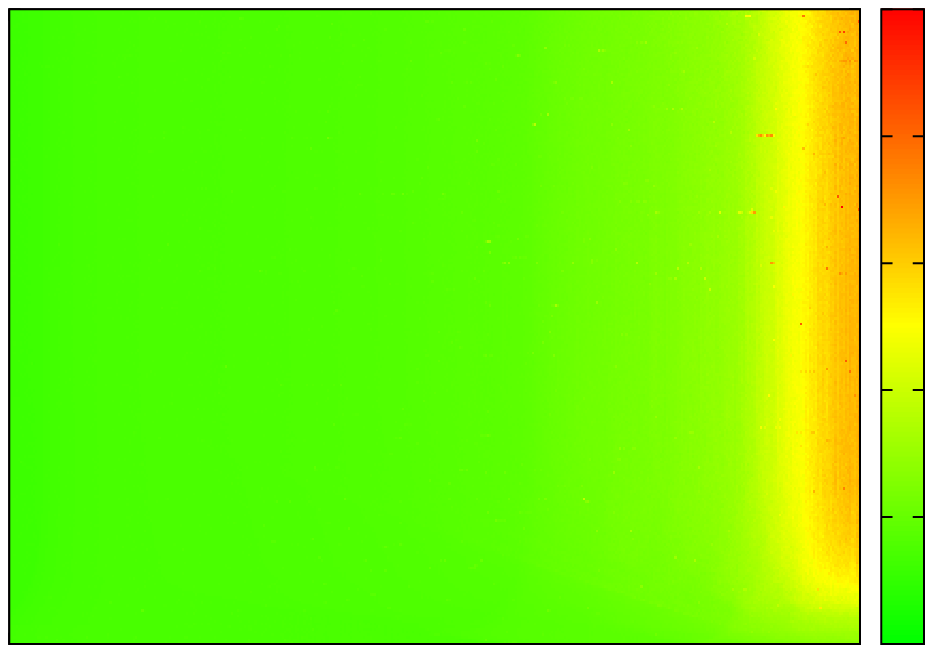 llx=0 lly=0 urx=360 ury=227 rwi=3600}
  \put(6369,4295){\makebox(0,0)[l]{\strut{}1.0}}%
  \put(6369,3564){\makebox(0,0)[l]{\strut{}0.8}}%
  \put(6369,2833){\makebox(0,0)[l]{\strut{}0.6}}%
  \put(6369,2102){\makebox(0,0)[l]{\strut{}0.4}}%
  \put(6369,1371){\makebox(0,0)[l]{\strut{}0.2}}%
  \put(6369,640){\makebox(0,0)[l]{\strut{}0.0}}%
  \put(3430,140){\makebox(0,0){\strut{}$\log_{10}(h/T)$}}%
  \put(160,2467){%
  \special{ps: gsave currentpoint currentpoint translate
630 rotate neg exch neg exch translate}%
  \makebox(0,0){\strut{}$\log_{10} (1 - e)$}%
  \special{ps: currentpoint grestore moveto}%
  }%
  \put(5881,440){\makebox(0,0){\strut{}$0$}}%
  \put(4249,440){\makebox(0,0){\strut{}$-1$}}%
  \put(2612,440){\makebox(0,0){\strut{}$-2$}}%
  \put(980,440){\makebox(0,0){\strut{}$-3$}}%
  \put(860,4295){\makebox(0,0)[r]{\strut{}$\mbox{\em}-8$}}%
  \put(860,3385){\makebox(0,0)[r]{\strut{}$\mbox{\em}-6$}}%
  \put(860,2468){\makebox(0,0)[r]{\strut{}$\mbox{\em}-4$}}%
  \put(860,1550){\makebox(0,0)[r]{\strut{}$\mbox{\em}-2$}}%
  \put(860,640){\makebox(0,0)[r]{\strut{}$\mbox{\em}0$}}%
\end{picture}%
\endgroup
 
\caption{The summary diagrams in the elliptic
case for {\tt universal.c} are plotted.  The top plot shows the relative energy error as a
function of eccentricity $e$ and stepsize $h$.  The color bar shows
the common logarithm of the magnitude.  The middle plot illustrates
bias in the calculation, plotting 1 if the energy error is positive,
-1 if the energy error is negative, and 0 for zero energy error.  The
bottom plot shows the computing time in microseconds per call to the Kepler stepper.}
\label{fig:fig1}
\end{figure*}

\begin{figure*}
\begingroup%
\makeatletter%
\newcommand{\GNUPLOTspecial}{%
  \@sanitize\catcode`\%=14\relax\special}%
\setlength{\unitlength}{0.0500bp}%
\begin{picture}(7200,4536)(0,0)%
  \special{psfile=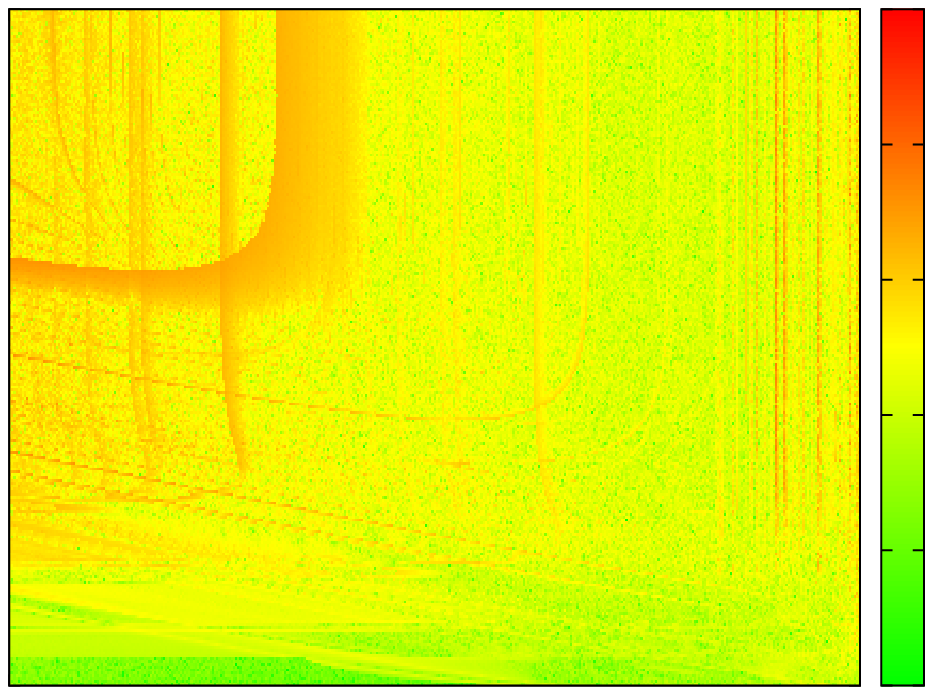 llx=0 lly=0 urx=360 ury=227 rwi=3600}
  \put(6369,4295){\makebox(0,0)[l]{\strut{}-6.0}}%
  \put(6369,3516){\makebox(0,0)[l]{\strut{}-8.0}}%
  \put(6369,2737){\makebox(0,0)[l]{\strut{}-10.0}}%
  \put(6369,1958){\makebox(0,0)[l]{\strut{}-12.0}}%
  \put(6369,1179){\makebox(0,0)[l]{\strut{}-14.0}}%
  \put(6369,400){\makebox(0,0)[l]{\strut{}-16.0}}%
  \put(160,2347){%
  \special{ps: gsave currentpoint currentpoint translate
630 rotate neg exch neg exch translate}%
  \makebox(0,0){\strut{}$\log_{10} (1 - e)$}%
  \special{ps: currentpoint grestore moveto}%
  }%
  \put(5881,200){\makebox(0,0){\strut{}$\ $}}%
  \put(4249,200){\makebox(0,0){\strut{}$\ $}}%
  \put(2612,200){\makebox(0,0){\strut{}$\ $}}%
  \put(980,200){\makebox(0,0){\strut{}$\ $}}%
  \put(860,4295){\makebox(0,0)[r]{\strut{}$\mbox{\em}-8$}}%
  \put(860,3325){\makebox(0,0)[r]{\strut{}$\mbox{\em}-6$}}%
  \put(860,2348){\makebox(0,0)[r]{\strut{}$\mbox{\em}-4$}}%
  \put(860,1370){\makebox(0,0)[r]{\strut{}$\mbox{\em}-2$}}%
  \put(860,400){\makebox(0,0)[r]{\strut{}$\mbox{\em}0$}}%
\end{picture}%
\endgroup
 
\vskip -20pt
\begingroup%
\makeatletter%
\newcommand{\GNUPLOTspecial}{%
  \@sanitize\catcode`\%=14\relax\special}%
\setlength{\unitlength}{0.0500bp}%
\begin{picture}(7200,4536)(0,0)%
  \special{psfile=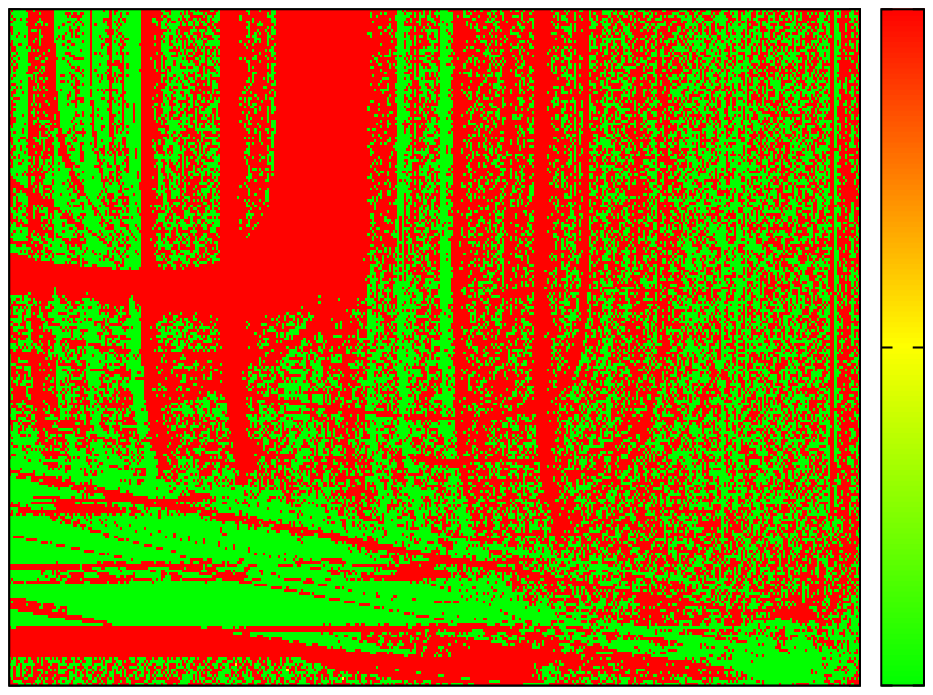 llx=0 lly=0 urx=360 ury=227 rwi=3600}
  \put(6369,4295){\makebox(0,0)[l]{\strut{}1.0}}%
  \put(6369,2347){\makebox(0,0)[l]{\strut{}0.0}}%
  \put(6369,400){\makebox(0,0)[l]{\strut{}-1.0}}%
  \put(160,2347){%
  \special{ps: gsave currentpoint currentpoint translate
630 rotate neg exch neg exch translate}%
  \makebox(0,0){\strut{}$\log_{10} (1 - e)$}%
  \special{ps: currentpoint grestore moveto}%
  }%
  \put(4249,200){\makebox(0,0){\strut{}$\ $}}%
  \put(2612,200){\makebox(0,0){\strut{}$\ $}}%
  \put(860,4295){\makebox(0,0)[r]{\strut{}$\mbox{\em}-8$}}%
  \put(860,3325){\makebox(0,0)[r]{\strut{}$\mbox{\em}-6$}}%
  \put(860,2348){\makebox(0,0)[r]{\strut{}$\mbox{\em}-4$}}%
  \put(860,1370){\makebox(0,0)[r]{\strut{}$\mbox{\em}-2$}}%
  \put(860,400){\makebox(0,0)[r]{\strut{}$\mbox{\em}0$}}%
\end{picture}%
\endgroup
 
\vskip -20pt
\begingroup%
\makeatletter%
\newcommand{\GNUPLOTspecial}{%
  \@sanitize\catcode`\%=14\relax\special}%
\setlength{\unitlength}{0.0500bp}%
\begin{picture}(7200,4536)(0,0)%
  \special{psfile=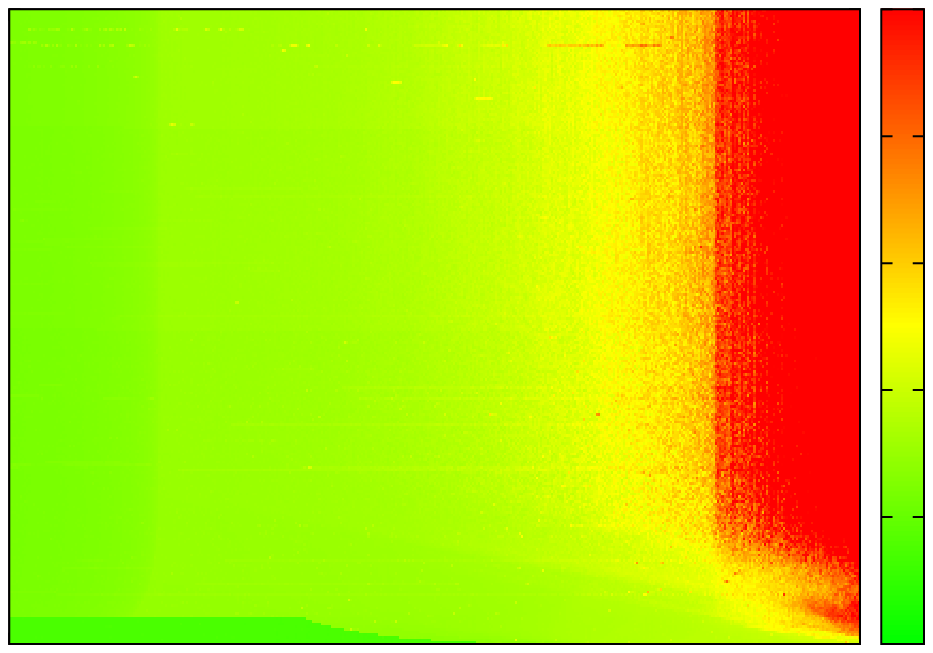 llx=0 lly=0 urx=360 ury=227 rwi=3600}
  \put(6369,4295){\makebox(0,0)[l]{\strut{}1.0}}%
  \put(6369,3564){\makebox(0,0)[l]{\strut{}0.8}}%
  \put(6369,2833){\makebox(0,0)[l]{\strut{}0.6}}%
  \put(6369,2102){\makebox(0,0)[l]{\strut{}0.4}}%
  \put(6369,1371){\makebox(0,0)[l]{\strut{}0.2}}%
  \put(6369,640){\makebox(0,0)[l]{\strut{}0.0}}%
  \put(3430,140){\makebox(0,0){\strut{}$\log_{10}(h/T)$}}%
  \put(160,2467){%
  \special{ps: gsave currentpoint currentpoint translate
630 rotate neg exch neg exch translate}%
  \makebox(0,0){\strut{}$\log_{10} (1 - e)$}%
  \special{ps: currentpoint grestore moveto}%
  }%
  \put(5881,440){\makebox(0,0){\strut{}$0$}}%
  \put(4249,440){\makebox(0,0){\strut{}$-1$}}%
  \put(2612,440){\makebox(0,0){\strut{}$-2$}}%
  \put(980,440){\makebox(0,0){\strut{}$-3$}}%
  \put(860,4295){\makebox(0,0)[r]{\strut{}$\mbox{\em}-8$}}%
  \put(860,3385){\makebox(0,0)[r]{\strut{}$\mbox{\em}-6$}}%
  \put(860,2468){\makebox(0,0)[r]{\strut{}$\mbox{\em}-4$}}%
  \put(860,1550){\makebox(0,0)[r]{\strut{}$\mbox{\em}-2$}}%
  \put(860,640){\makebox(0,0)[r]{\strut{}$\mbox{\em}0$}}%
\end{picture}%
\endgroup
 
\caption{The summary diagrams in the elliptic
case for {\tt drift\_one.c} are plotted.  The details are the same as in Fig.~\ref{fig:fig1}.}
\label{fig:fig2}
\end{figure*}

\begin{figure*}
\begingroup%
\makeatletter%
\newcommand{\GNUPLOTspecial}{%
  \@sanitize\catcode`\%=14\relax\special}%
\setlength{\unitlength}{0.0500bp}%
\begin{picture}(7200,4536)(0,0)%
  \special{psfile=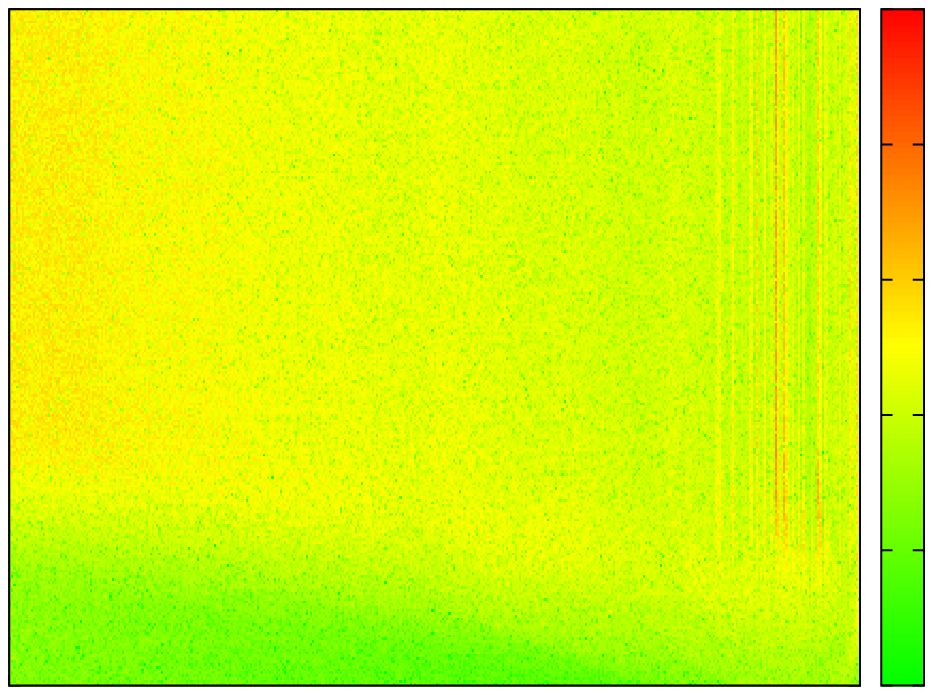 llx=0 lly=0 urx=360 ury=227 rwi=3600}
  \put(6369,4295){\makebox(0,0)[l]{\strut{}-6.0}}%
  \put(6369,3516){\makebox(0,0)[l]{\strut{}-8.0}}%
  \put(6369,2737){\makebox(0,0)[l]{\strut{}-10.0}}%
  \put(6369,1958){\makebox(0,0)[l]{\strut{}-12.0}}%
  \put(6369,1179){\makebox(0,0)[l]{\strut{}-14.0}}%
  \put(6369,400){\makebox(0,0)[l]{\strut{}-16.0}}%
  \put(160,2347){%
  \special{ps: gsave currentpoint currentpoint translate
630 rotate neg exch neg exch translate}%
  \makebox(0,0){\strut{}$\log_{10} (1 - e)$}%
  \special{ps: currentpoint grestore moveto}%
  }%
  \put(5881,200){\makebox(0,0){\strut{}$\ $}}%
  \put(4249,200){\makebox(0,0){\strut{}$\ $}}%
  \put(2612,200){\makebox(0,0){\strut{}$\ $}}%
  \put(980,200){\makebox(0,0){\strut{}$\ $}}%
  \put(860,4295){\makebox(0,0)[r]{\strut{}$\mbox{\em}-8$}}%
  \put(860,3325){\makebox(0,0)[r]{\strut{}$\mbox{\em}-6$}}%
  \put(860,2348){\makebox(0,0)[r]{\strut{}$\mbox{\em}-4$}}%
  \put(860,1370){\makebox(0,0)[r]{\strut{}$\mbox{\em}-2$}}%
  \put(860,400){\makebox(0,0)[r]{\strut{}$\mbox{\em}0$}}%
\end{picture}%
\endgroup
 
\vskip -20pt
\begingroup%
\makeatletter%
\newcommand{\GNUPLOTspecial}{%
  \@sanitize\catcode`\%=14\relax\special}%
\setlength{\unitlength}{0.0500bp}%
\begin{picture}(7200,4536)(0,0)%
  \special{psfile=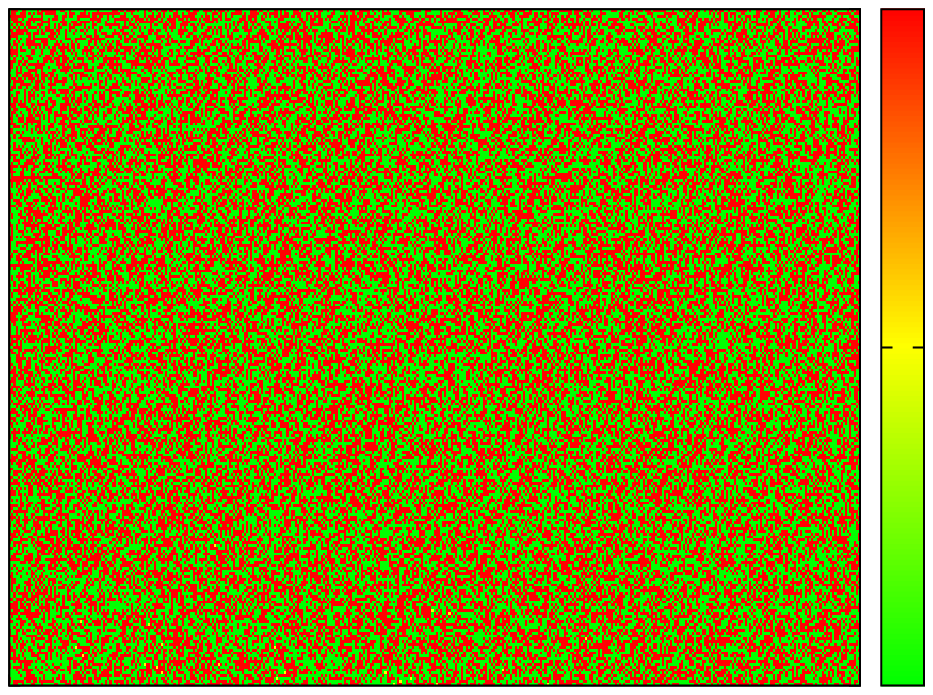 llx=0 lly=0 urx=360 ury=227 rwi=3600}
  \put(6369,2347){\makebox(0,0)[l]{\strut{}0.0}}%
  \put(160,2347){%
  \special{ps: gsave currentpoint currentpoint translate
630 rotate neg exch neg exch translate}%
  \makebox(0,0){\strut{}$\log_{10} (1 - e)$}%
  \special{ps: currentpoint grestore moveto}%
  }%
  \put(5881,200){\makebox(0,0){\strut{}$\ $}}%
  \put(4249,200){\makebox(0,0){\strut{}$\ $}}%
  \put(2612,200){\makebox(0,0){\strut{}$\ $}}%
  \put(980,200){\makebox(0,0){\strut{}$\ $}}%
  \put(860,4295){\makebox(0,0)[r]{\strut{}$\mbox{\em}-8$}}%
  \put(860,3325){\makebox(0,0)[r]{\strut{}$\mbox{\em}-6$}}%
  \put(860,2348){\makebox(0,0)[r]{\strut{}$\mbox{\em}-4$}}%
  \put(860,1370){\makebox(0,0)[r]{\strut{}$\mbox{\em}-2$}}%
  \put(860,400){\makebox(0,0)[r]{\strut{}$\mbox{\em}0$}}%
\end{picture}%
\endgroup
 
\vskip -20pt
\begingroup%
\makeatletter%
\newcommand{\GNUPLOTspecial}{%
  \@sanitize\catcode`\%=14\relax\special}%
\setlength{\unitlength}{0.0500bp}%
\begin{picture}(7200,4536)(0,0)%
  \special{psfile=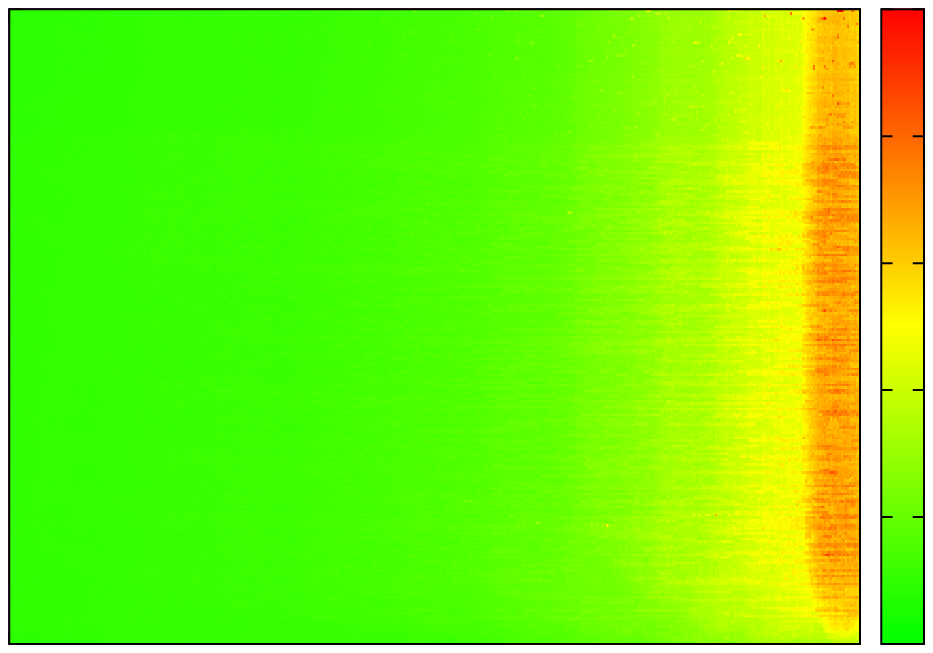 llx=0 lly=0 urx=360 ury=227 rwi=3600}
  \put(6369,4295){\makebox(0,0)[l]{\strut{}1.0}}%
  \put(6369,3564){\makebox(0,0)[l]{\strut{}0.8}}%
  \put(6369,2833){\makebox(0,0)[l]{\strut{}0.6}}%
  \put(6369,2102){\makebox(0,0)[l]{\strut{}0.4}}%
  \put(6369,1371){\makebox(0,0)[l]{\strut{}0.2}}%
  \put(6369,640){\makebox(0,0)[l]{\strut{}0.0}}%
  \put(3430,140){\makebox(0,0){\strut{}$\log_{10}(h/T)$}}%
  \put(160,2467){%
  \special{ps: gsave currentpoint currentpoint translate
630 rotate neg exch neg exch translate}%
  \makebox(0,0){\strut{}$\log_{10} (1 - e)$}%
  \special{ps: currentpoint grestore moveto}%
  }%
  \put(5881,440){\makebox(0,0){\strut{}$0$}}%
  \put(4249,440){\makebox(0,0){\strut{}$-1$}}%
  \put(2612,440){\makebox(0,0){\strut{}$-2$}}%
  \put(980,440){\makebox(0,0){\strut{}$-3$}}%
  \put(860,4295){\makebox(0,0)[r]{\strut{}$\mbox{\em}-8$}}%
  \put(860,3385){\makebox(0,0)[r]{\strut{}$\mbox{\em}-6$}}%
  \put(860,2468){\makebox(0,0)[r]{\strut{}$\mbox{\em}-4$}}%
  \put(860,1550){\makebox(0,0)[r]{\strut{}$\mbox{\em}-2$}}%
  \put(860,640){\makebox(0,0)[r]{\strut{}$\mbox{\em}0$}}%
\end{picture}%
\endgroup
 
\caption{The summary diagrams in the elliptic
case for the Kepler solver in WHFast are plotted.  The details are the same as in Fig.~\ref{fig:fig1}.}
\label{fig:fig3}
\end{figure*}

\begin{figure*}
\begingroup%
\makeatletter%
\newcommand{\GNUPLOTspecial}{%
  \@sanitize\catcode`\%=14\relax\special}%
\setlength{\unitlength}{0.0500bp}%
\begin{picture}(7200,4536)(0,0)%
  \special{psfile=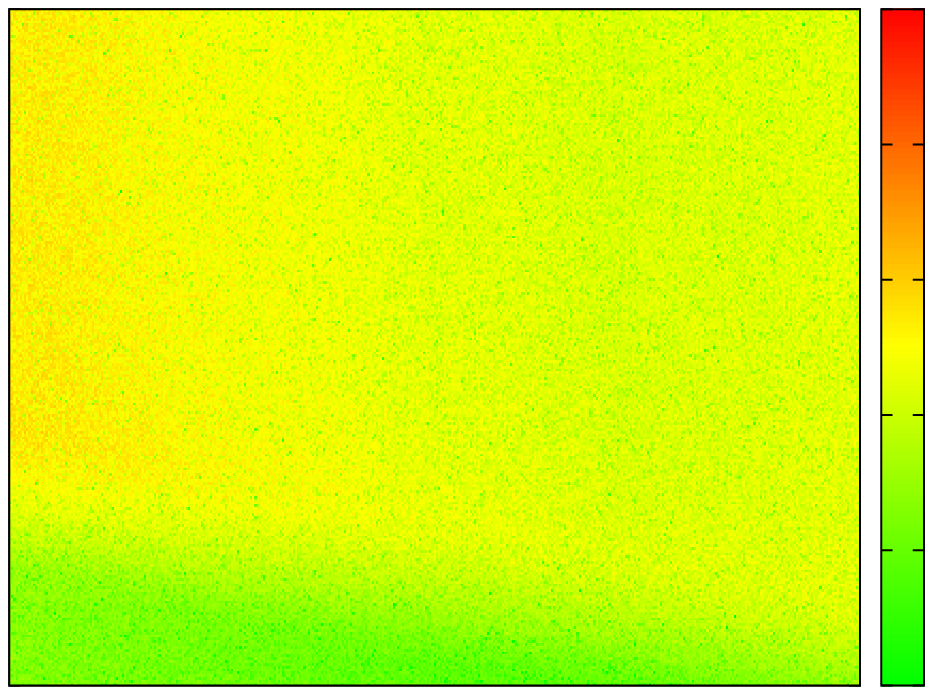 llx=0 lly=0 urx=360 ury=227 rwi=3600}
  \put(6369,4295){\makebox(0,0)[l]{\strut{}-6.0}}%
  \put(6369,3516){\makebox(0,0)[l]{\strut{}-8.0}}%
  \put(6369,2737){\makebox(0,0)[l]{\strut{}-10.0}}%
  \put(6369,1958){\makebox(0,0)[l]{\strut{}-12.0}}%
  \put(6369,1179){\makebox(0,0)[l]{\strut{}-14.0}}%
  \put(6369,400){\makebox(0,0)[l]{\strut{}-16.0}}%
  \put(160,2347){%
  \special{ps: gsave currentpoint currentpoint translate
630 rotate neg exch neg exch translate}%
  \makebox(0,0){\strut{}$\log_{10} (e - 1)$}%
  \special{ps: currentpoint grestore moveto}%
  }%
  \put(4249,200){\makebox(0,0){\strut{}$\ $}}%
  \put(2612,200){\makebox(0,0){\strut{}$\ $}}%
  \put(860,4295){\makebox(0,0)[r]{\strut{}$\mbox{\em}-8$}}%
  \put(860,3325){\makebox(0,0)[r]{\strut{}$\mbox{\em}-6$}}%
  \put(860,2348){\makebox(0,0)[r]{\strut{}$\mbox{\em}-4$}}%
  \put(860,1370){\makebox(0,0)[r]{\strut{}$\mbox{\em}-2$}}%
  \put(860,400){\makebox(0,0)[r]{\strut{}$\mbox{\em}0$}}%
\end{picture}%
\endgroup
 
\vskip -20pt
\begingroup%
\makeatletter%
\newcommand{\GNUPLOTspecial}{%
  \@sanitize\catcode`\%=14\relax\special}%
\setlength{\unitlength}{0.0500bp}%
\begin{picture}(7200,4536)(0,0)%
  \special{psfile=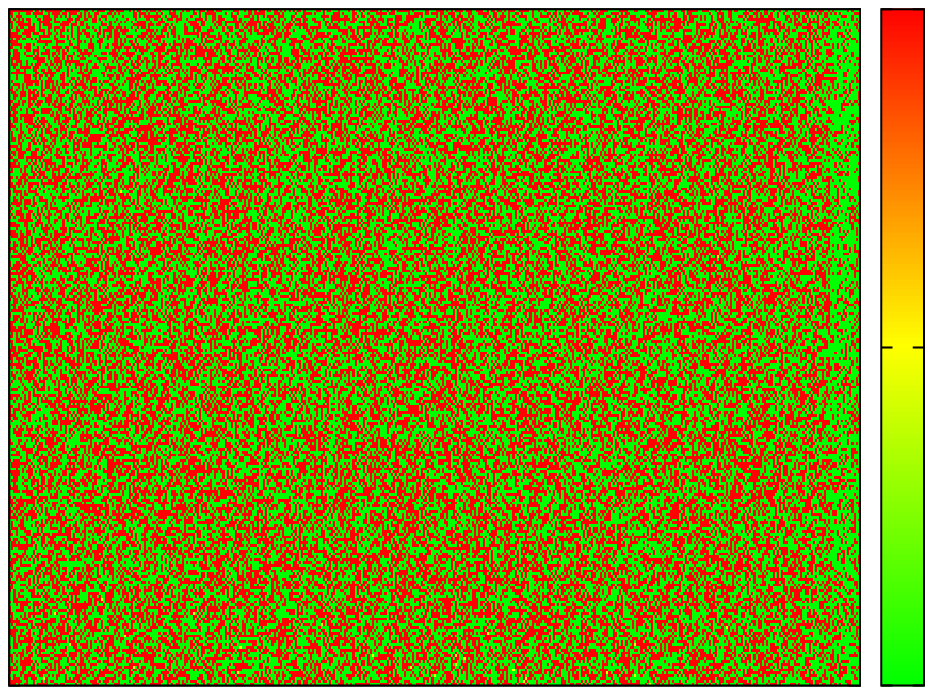 llx=0 lly=0 urx=360 ury=227 rwi=3600}
  \put(6369,4295){\makebox(0,0)[l]{\strut{}1.0}}%
  \put(6369,2347){\makebox(0,0)[l]{\strut{}0.0}}%
  \put(6369,400){\makebox(0,0)[l]{\strut{}-1.0}}%
  \put(160,2347){%
  \special{ps: gsave currentpoint currentpoint translate
630 rotate neg exch neg exch translate}%
  \makebox(0,0){\strut{}$\log_{10} (e - 1)$}%
  \special{ps: currentpoint grestore moveto}%
  }%
  \put(5881,200){\makebox(0,0){\strut{}$\ $}}%
  \put(4249,200){\makebox(0,0){\strut{}$\ $}}%
  \put(2612,200){\makebox(0,0){\strut{}$\ $}}%
  \put(980,200){\makebox(0,0){\strut{}$\ $}}%
  \put(860,4295){\makebox(0,0)[r]{\strut{}$\mbox{\em}-8$}}%
  \put(860,3325){\makebox(0,0)[r]{\strut{}$\mbox{\em}-6$}}%
  \put(860,2348){\makebox(0,0)[r]{\strut{}$\mbox{\em}-4$}}%
  \put(860,1370){\makebox(0,0)[r]{\strut{}$\mbox{\em}-2$}}%
  \put(860,400){\makebox(0,0)[r]{\strut{}$\mbox{\em}0$}}%
\end{picture}%
\endgroup
 
\vskip -20pt
\begingroup%
\makeatletter%
\newcommand{\GNUPLOTspecial}{%
  \@sanitize\catcode`\%=14\relax\special}%
\setlength{\unitlength}{0.0500bp}%
\begin{picture}(7200,4536)(0,0)%
  \special{psfile=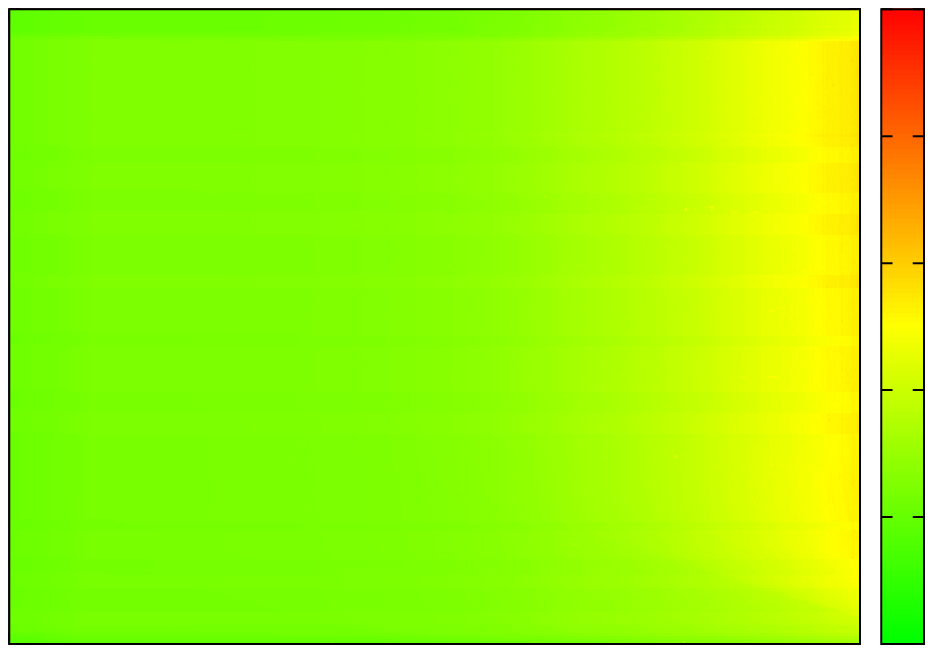 llx=0 lly=0 urx=360 ury=227 rwi=3600}
  \put(6369,4295){\makebox(0,0)[l]{\strut{}1.0}}%
  \put(6369,3564){\makebox(0,0)[l]{\strut{}0.8}}%
  \put(6369,2833){\makebox(0,0)[l]{\strut{}0.6}}%
  \put(6369,2102){\makebox(0,0)[l]{\strut{}0.4}}%
  \put(6369,1371){\makebox(0,0)[l]{\strut{}0.2}}%
  \put(6369,640){\makebox(0,0)[l]{\strut{}0.0}}%
  \put(3430,140){\makebox(0,0){\strut{}$\log_{10}(h/T)$}}%
  \put(160,2467){%
  \special{ps: gsave currentpoint currentpoint translate
630 rotate neg exch neg exch translate}%
  \makebox(0,0){\strut{}$\log_{10} (e - 1)$}%
  \special{ps: currentpoint grestore moveto}%
  }%
  \put(5881,440){\makebox(0,0){\strut{}$0$}}%
  \put(4249,440){\makebox(0,0){\strut{}$-1$}}%
  \put(2612,440){\makebox(0,0){\strut{}$-2$}}%
  \put(980,440){\makebox(0,0){\strut{}$-3$}}%
  \put(860,4295){\makebox(0,0)[r]{\strut{}$\mbox{\em}-8$}}%
  \put(860,3385){\makebox(0,0)[r]{\strut{}$\mbox{\em}-6$}}%
  \put(860,2468){\makebox(0,0)[r]{\strut{}$\mbox{\em}-4$}}%
  \put(860,1550){\makebox(0,0)[r]{\strut{}$\mbox{\em}-2$}}%
  \put(860,640){\makebox(0,0)[r]{\strut{}$\mbox{\em}0$}}%
\end{picture}%
\endgroup
 
\caption{The summary diagrams in the hyperbolic
case for {\tt universal.c} are plotted.  The details are the same as in Fig.~\ref{fig:fig1}.}
\label{fig:fig4}
\end{figure*}

\begin{figure*}
\begingroup%
\makeatletter%
\newcommand{\GNUPLOTspecial}{%
  \@sanitize\catcode`\%=14\relax\special}%
\setlength{\unitlength}{0.0500bp}%
\begin{picture}(7200,4536)(0,0)%
  \special{psfile=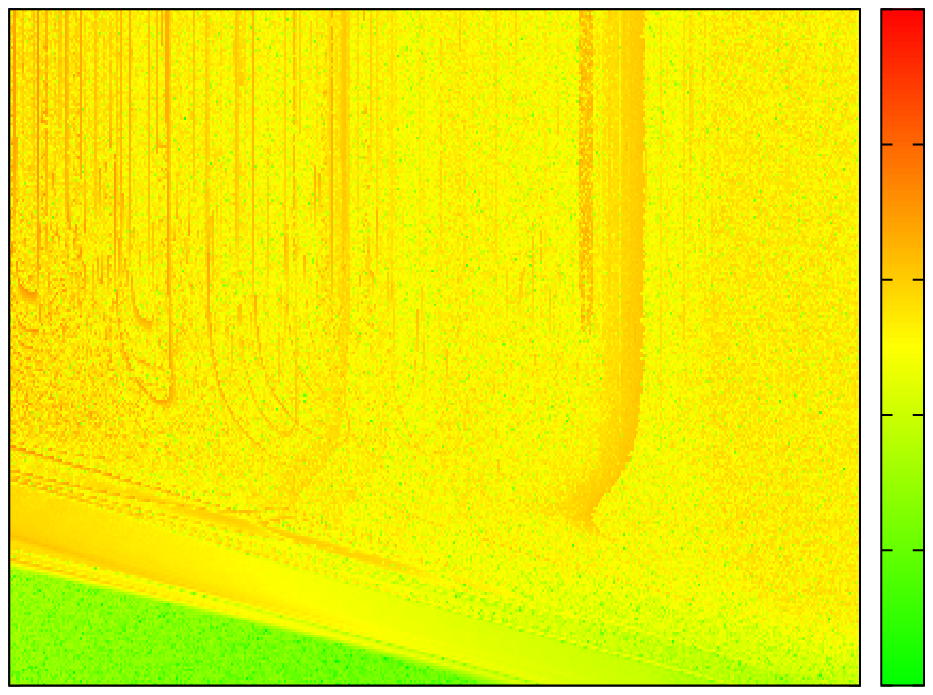 llx=0 lly=0 urx=360 ury=227 rwi=3600}
  \put(6369,4295){\makebox(0,0)[l]{\strut{}-6.0}}%
  \put(6369,3516){\makebox(0,0)[l]{\strut{}-8.0}}%
  \put(6369,2737){\makebox(0,0)[l]{\strut{}-10.0}}%
  \put(6369,1958){\makebox(0,0)[l]{\strut{}-12.0}}%
  \put(6369,1179){\makebox(0,0)[l]{\strut{}-14.0}}%
  \put(6369,400){\makebox(0,0)[l]{\strut{}-16.0}}%
  \put(160,2347){%
  \special{ps: gsave currentpoint currentpoint translate
630 rotate neg exch neg exch translate}%
  \makebox(0,0){\strut{}$\log_{10} (e - 1)$}%
  \special{ps: currentpoint grestore moveto}%
  }%
  \put(4249,200){\makebox(0,0){\strut{}$\ $}}%
  \put(2612,200){\makebox(0,0){\strut{}$\ $}}%
  \put(860,4295){\makebox(0,0)[r]{\strut{}$\mbox{\em}-8$}}%
  \put(860,3325){\makebox(0,0)[r]{\strut{}$\mbox{\em}-6$}}%
  \put(860,2348){\makebox(0,0)[r]{\strut{}$\mbox{\em}-4$}}%
  \put(860,1370){\makebox(0,0)[r]{\strut{}$\mbox{\em}-2$}}%
  \put(860,400){\makebox(0,0)[r]{\strut{}$\mbox{\em}0$}}%
\end{picture}%
\endgroup
 
\vskip -20pt
\begingroup%
\makeatletter%
\newcommand{\GNUPLOTspecial}{%
  \@sanitize\catcode`\%=14\relax\special}%
\setlength{\unitlength}{0.0500bp}%
\begin{picture}(7200,4536)(0,0)%
  \special{psfile=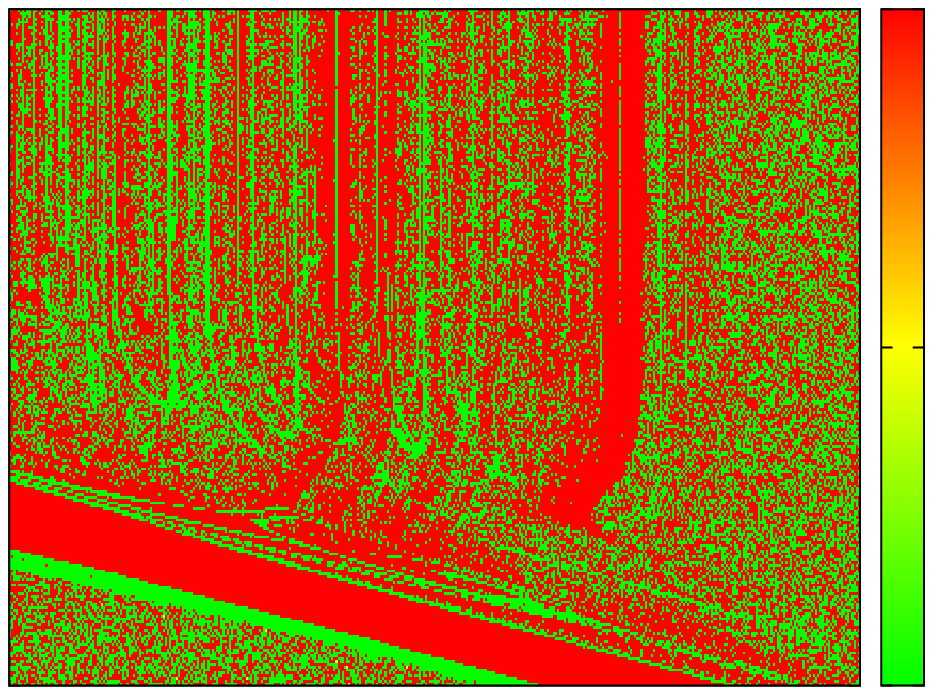 llx=0 lly=0 urx=360 ury=227 rwi=3600}
  \put(6369,4295){\makebox(0,0)[l]{\strut{}1.0}}%
  \put(6369,2347){\makebox(0,0)[l]{\strut{}0.0}}%
  \put(6369,400){\makebox(0,0)[l]{\strut{}-1.0}}%
  \put(160,2347){%
  \special{ps: gsave currentpoint currentpoint translate
630 rotate neg exch neg exch translate}%
  \makebox(0,0){\strut{}$\log_{10} (e - 1)$}%
  \special{ps: currentpoint grestore moveto}%
  }%
  \put(4249,200){\makebox(0,0){\strut{}$\ $}}%
  \put(2612,200){\makebox(0,0){\strut{}$\ $}}%
  \put(860,4295){\makebox(0,0)[r]{\strut{}$\mbox{\em}-8$}}%
  \put(860,3325){\makebox(0,0)[r]{\strut{}$\mbox{\em}-6$}}%
  \put(860,2348){\makebox(0,0)[r]{\strut{}$\mbox{\em}-4$}}%
  \put(860,1370){\makebox(0,0)[r]{\strut{}$\mbox{\em}-2$}}%
  \put(860,400){\makebox(0,0)[r]{\strut{}$\mbox{\em}0$}}%
\end{picture}%
\endgroup
 
\vskip -20pt
\begingroup%
\makeatletter%
\newcommand{\GNUPLOTspecial}{%
  \@sanitize\catcode`\%=14\relax\special}%
\setlength{\unitlength}{0.0500bp}%
\begin{picture}(7200,4536)(0,0)%
  \special{psfile=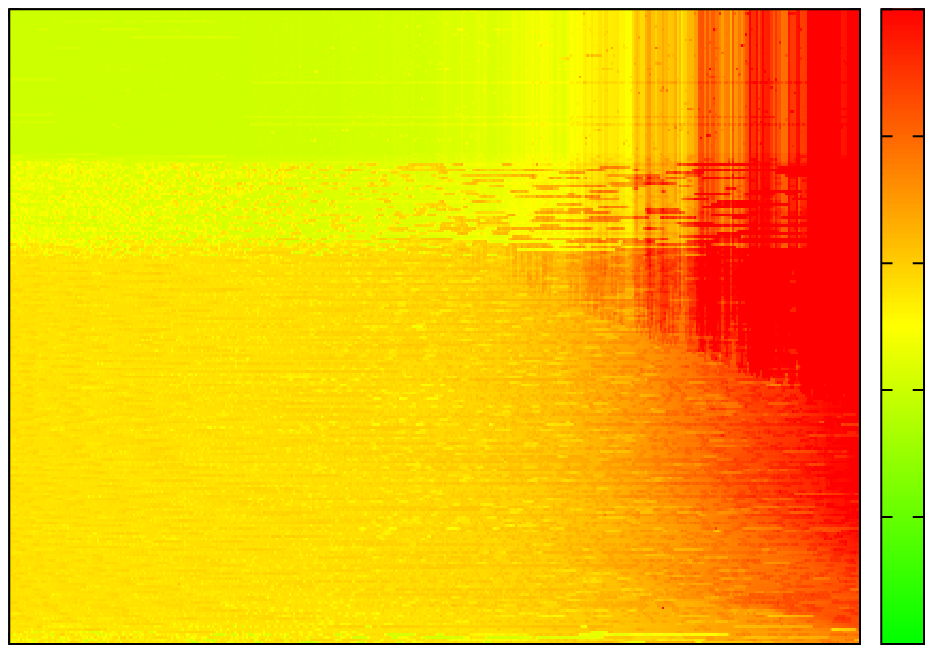 llx=0 lly=0 urx=360 ury=227 rwi=3600}
  \put(6369,4295){\makebox(0,0)[l]{\strut{}1.0}}%
  \put(6369,3564){\makebox(0,0)[l]{\strut{}0.8}}%
  \put(6369,2833){\makebox(0,0)[l]{\strut{}0.6}}%
  \put(6369,2102){\makebox(0,0)[l]{\strut{}0.4}}%
  \put(6369,1371){\makebox(0,0)[l]{\strut{}0.2}}%
  \put(6369,640){\makebox(0,0)[l]{\strut{}0.0}}%
  \put(3430,140){\makebox(0,0){\strut{}$\log_{10}(h/T)$}}%
  \put(160,2467){%
  \special{ps: gsave currentpoint currentpoint translate
630 rotate neg exch neg exch translate}%
  \makebox(0,0){\strut{}$\log_{10} (e - 1)$}%
  \special{ps: currentpoint grestore moveto}%
  }%
  \put(5881,440){\makebox(0,0){\strut{}$0$}}%
  \put(4249,440){\makebox(0,0){\strut{}$-1$}}%
  \put(2612,440){\makebox(0,0){\strut{}$-2$}}%
  \put(980,440){\makebox(0,0){\strut{}$-3$}}%
  \put(860,4295){\makebox(0,0)[r]{\strut{}$\mbox{\em}-8$}}%
  \put(860,3385){\makebox(0,0)[r]{\strut{}$\mbox{\em}-6$}}%
  \put(860,2468){\makebox(0,0)[r]{\strut{}$\mbox{\em}-4$}}%
  \put(860,1550){\makebox(0,0)[r]{\strut{}$\mbox{\em}-2$}}%
  \put(860,640){\makebox(0,0)[r]{\strut{}$\mbox{\em}0$}}%
\end{picture}%
\endgroup
 
\caption{The summary diagrams in the hyperbolic
case for {\tt drift\_one.c} are plotted.  The details are the same as in Fig.~\ref{fig:fig1}.}
\label{fig:fig5}
\end{figure*}

In Fig.~\ref{fig:histogram-hyperbolic}, histograms of the relative
energy error in the hyperbolic case for {\tt drift\_one.c} and {\tt
universal.c} are shown.  We see that {\tt drift\_one.c} is typically
less accurate than {\tt universal.c}.  Quantitatively, the 
averages of the common logarithm of the absolute value of the relative
error are: -11.72 and -11.03, for {\tt universal.c} and {\tt
drift\_one.c}, respectively.  The error of {\tt universal.c} is, on
average, about a factor of 5 smaller than the error in {\tt
drift\_one.c}.

\section{Summary}

We have developed and tested a universal variable solver for the
Kepler initial value problem.  The method eschews the use of Stumpff
series in favor of ordinary trigonometric functions.  We have been
careful to make sure that all expressions are numerically well
defined.  We find that our program performs better, in terms of
accuracy, bias, and speed, than a C version of a widely used Kepler
solver, which uses the Stumpff series.  Our code is freely available;
contact the authors.

\section*{Acknowledgements}

We thank Hanno Rein, Daniel Tamayo, and Edmund Bertschinger for helpful discussions.
DMH acknowledges support by a NSF Graduate Research Fellowship under
Grant No. 1122374.




\bibliographystyle{mn2e}

\bibliography{wisdom-mnras2}


\bsp	
\label{lastpage}
\end{document}